\documentclass[preprint,12pt]{elsarticle}

\usepackage{amssymb}
\usepackage{amsmath}
\usepackage{algorithm} 
\usepackage{algpseudocode} 
\usepackage{url}
\usepackage{tabularx}

\journal{Nuclear Physics B}

\begin{document}

\begin{frontmatter}



\title{The-Bodega: A Matlab Toolbox for Biologically Dynamic Microbubble Simulations on Realistic Hemodynamic Microvascular Graphs}


\author[label1]{Stephen Alexander Lee}
\author[label1]{Alexis Leconte} 
\author[label1]{Alice Wu}
\author[label1]{Jonathan Poree}
\author[label1]{Maxence Laplante-Berthier}
\author[label1]{Simon Desrocher} 
\author[label1]{Pierre-Olivier Bouchard} %% Author name
\author[label2]{Joshua Kinugasa}
\author[label3]{Samuel Mihelic}
\author[label3]{Andreas Linninger}
\author[label1,label4]{Jean Provost}

\affiliation[label1]{Department={Department of Engineering Physics},
            organization={Polytechnique Montreal},%Department and Organization
            addressline={2500 Chemin Polytechnique}, 
            city={Montreal},
            postcode={H3T 0A3}, 
            state={Quebec},
            country={Canada}}
\affiliation[label2]{Department={Department of Biomedical Engineering},
            organization={Chiba University},%Department and Organization
            addressline={1-33 Yayoicho}, 
            city={Inage Ward},
            postcode={263-8522}, 
            state={Chiba},
            country={Japan}}
\affiliation[label3]{Department={Department of Biomedical Engineering},
            organization={University of Illinois Chicago},%Department and Organization            
            addressline={1200 W Harrison St}, 
            city={Chicago},
            postcode={60607}, 
            state={Illinois},
            country={USA}}
\affiliation[label4]{organization={Montreal Heart Institute},%Department and Organization            
            addressline={5000 Rue Bélanger}, 
            city={Montreal},
            postcode={H1T 1C8}, 
            state={Quebec},
            country={Canada}}
%% Abstract
\begin{abstract}
The-Bodega is a \textsc{Matlab}-based toolbox for simulating ground-truth datasets for Ultrasound Localization Microscopy (ULM)—a super resolution imaging technique that resolves microvessels by systematically tracking microbubbles flowing through the microvasculature. The-Bodega enables open-source simulation of stochastic microbubble dynamics through anatomically complex vascular graphs and features a quasi-automated pipeline for generating ground-truth ultrasound data from simple vascular inputs. It incorporates sequential Monte Carlo simulations augmented with Poiseuille flow distributions and dynamic pulsatile flow. A key novelty of our framework is its flexibility to accommodate arbitrary vascular architectures and benchmark common ULM algorithms, such as Fourier Ring Correlation and Singular Value Decomposition (SVD) spatiotemporal filtering, on realistic hemodynamic digital phantoms. The-Bodega supports consistent microbubble-to-ultrasound simulations across domains ranging from mouse brains to human hearts and automatically leverages available CPU/GPU parallelization to improve computational efficiency. We demonstrate its versatility in applications including image quality assessment, motion artifact analysis, and the simulation of novel ULM modalities, such as capillary imaging, myocardial reconstruction under beating heart motion, and simulating neurovascular evoked responses.
\end{abstract}

\begin{graphicalabstract}
\end{graphicalabstract}


%% Keywords

\end{frontmatter}



\section{Introduction}
\label{sec:intro}

All organs in the human body are significantly perfused by an intricate network of blood vessels responsible for transporting nutrients and oxygen. Consequently, disruptions or abnormalities in the vascular system can lead to downstream deficits and pathological conditions. Cardiovascular diseases (CVD) have been the leading cause of death in the United States since 1921 \cite{doi:10.1161/CIR.0000000000001209}. Moreover, vascular abnormalities have been well documented in the brain, with increasing evidence linking vascular dysfunction to the progression of Alzheimer's disease, dementia, and stroke \cite{eisenmenger2023vascular}. Compounding this issue, disparities in socioeconomic, environmental, and psychosocial factors are strongly associated with the onset and mechanisms of CVD. These realities underscore the urgent need for high-throughput monitoring and treatment strategies that are tailored to the individual.

Among available medical imaging technologies, Doppler echocardiography is widely used to assess blood flow anomalies \cite{nishimura1985doppler}. However, despite high frame rates, traditional echocardiography is limited by poor skull penetration and insufficient resolution to measure small, low-flow vessels due to diffraction constraints and trade-offs between penetration depth and spatial resolution. Ultrasound localization microscopy (ULM), a super-resolution ultrasound technique inspired by optical localization microscopy, tracks distinguishable microbubbles as they flow through the microvasculature \cite{couture2018ultrasound}. The aggregation of thousands of these microbubble trajectories enables the reconstruction of arteries and veins with resolutions down to tens of micrometers---even through the skull. Despite the remarkable capabilities of ULM for imaging microvasculature in organs such as the brain \cite{hingot2021measuring}, heart \cite{yan2024transthoracic}, and kidneys \cite{denis2023sensing}, its functional and clinical value remains under active investigation and requires further validation.

Despite advances in imaging technologies, pharmaceuticals, and medical devices, conventional diagnostics and therapeutics still largely rely on empirical evidence and the aggregation of results from clinical trials across populations. While such studies may have the statistical power to determine outcomes for groups of patients with similar pathologies, they often lack the granularity to enable accurate prognosis at the individual level. Recent advances in computational cardiac simulations have demonstrated the feasibility of alternative frameworks that integrate patient-specific datasets to provide individualized diagnostic, therapeutic, and prognostic insights \cite{niederer2019computational}. Ultrasound localization microscopy may play a crucial role in this direction by offering patient-specific angiograms, serving as unique input for personalized models. However, to realize this potential, significant development is needed in ULM technology itself to enable its application in precision medicine. Currently, ULM still depends on empirical data to assess image quality, revealing a critical gap in computational tools capable of generating ground truth datasets and benchmarking essential algorithms under realistic microvascular flow conditions.

Several simulation packages have been developed to benchmark specific aspects of ultrasound localization microscopy, including the PALA toolkit \cite{heiles2022performance}, BUFF \cite{lerendegui2022bubble}, and nonlinear ultrasound simulators such as PROTEUS \cite{blanken2024proteus,11002619}. PALA has significantly advanced the field by introducing benchmarking metrics for vessel structure and providing both simple simulated and \textit{in vivo} datasets, with a particular strength in anatomical imaging applications. BUFF and PROTEUS have further enriched the simulation landscape by modeling microbubble interactions and nonlinear contrast mechanisms, enabling comprehensive studies of advanced imaging modes. These simulators have made valuable contributions to ULM development, each designed to address specific research objectives within their respective domains. Building upon these foundations, there remains an opportunity to develop simulation frameworks that emphasize modularity and adaptability for broader ULM workflows, particularly in anatomically realistic contexts with comprehensive functional validation. For \textit{in vivo} velocity measurements and downstream functional metrics to achieve scientific veracity \cite{stevenson2014does}, simulation frameworks benefit from incorporating high-fidelity, anatomically realistic models that account for common imaging artifacts such as clutter and motion. This approach complements existing tools by providing additional validation capabilities for microbubble tracking algorithms and enabling systematic investigation of the fundamental performance characteristics of ULM-derived functional metrics.

To address these challenges, we introduce The-Bodega (available at \url{to-be-published}%\url{https://github.com/ProvostUltrasoundLab/the-bodega}
), a fully open-source, modular simulation toolbox for modeling microbubble dynamics on anatomically and hemodynamically realistic vascular graph networks. Building on our prior work \cite{belgharbi2023anatomically}, The-Bodega simulates microbubble trajectories across realistic angioarchitectures of the mouse brain (both half- and whole-brain models) and human heart. The toolbox is structured around object-oriented MATLAB classes to promote usability, flexibility, and modularity. The-Bodega supports ultrasound data generation using a linear SIMUS-based implementation \cite{garcia2024simus3} and can be easily integrated with other simulators such as k-Wave \cite{treeby2010k} or FULLWAVE \cite{pinton2021fullwave}. It simulates complete datasets reflective of realistic ULM acquisitions over minutes, incorporating physiological clutter from the mouse skull and cardiac motion in the human heart. A key advantage of our framework is the propagation of \textit{a priori} ground truth information throughout the simulation pipeline, facilitating direct comparison with estimated outputs. These datasets are openly available and designed to accelerate the development, testing, and benchmarking of novel ULM and deep learning algorithms within the research community.

%% Labels are used to cross-reference an item using \ref command.

%% Use \subsection commands to start a subsection.
\section{Methods}
\label{Methods}

\begin{figure}[h]
    \centering
    \includegraphics[width=0.9\linewidth]{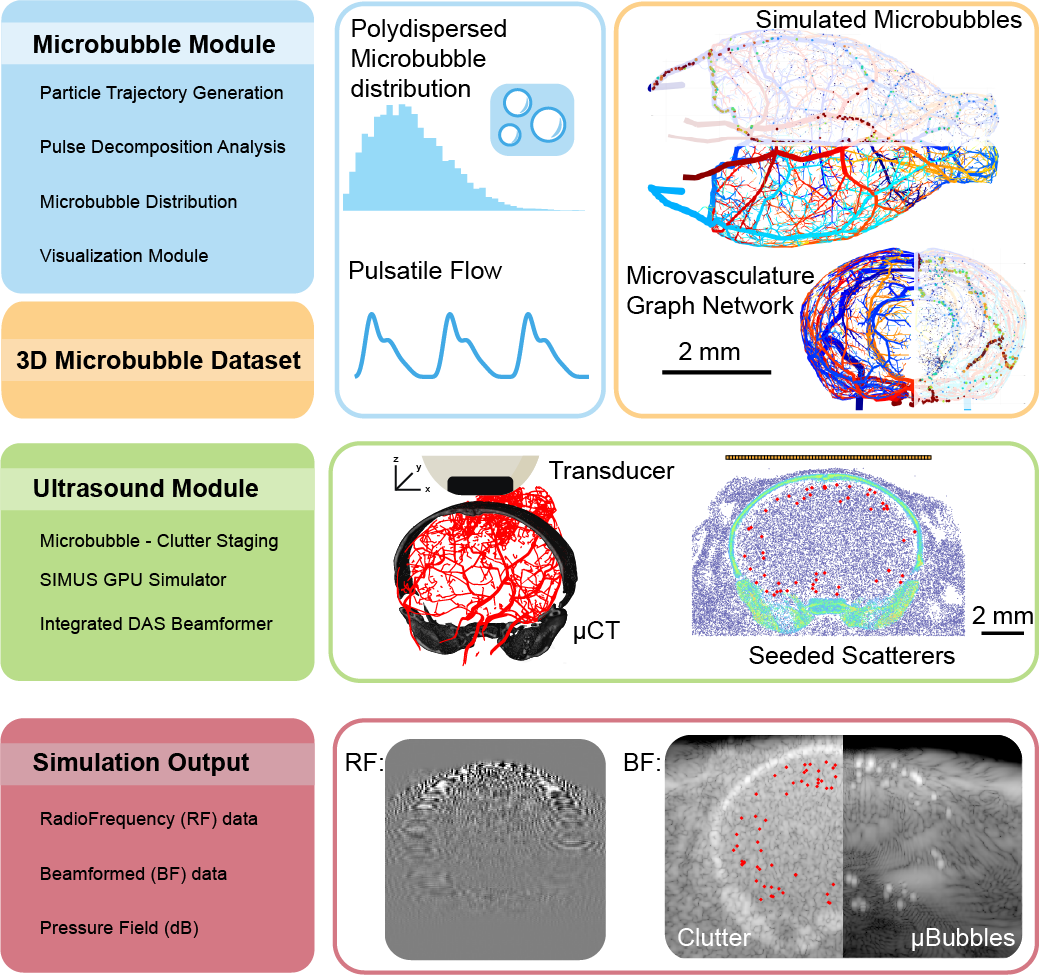}
    \caption{Pipeline adopted by The-Bodega to build ultrasound localization microscopy data from anatomically realistic 3D microvascular graphs.}
    \label{fig:1}
\end{figure}

The general pipeline of The-Bodega is illustrated in Figure~\ref{fig:1}. The process begins with microvascular directed graph networks composed of edges and nodes, each annotated with key physiological parameters such as blood flow velocity, vessel diameter, and pulse pressure. These networks serve as the structural scaffold for the microbubble simulation module. This module outputs a dataset by randomly sampling microbubble trajectories, stored in HDF5 format, which include ground truth information on microbubble position, velocity, pulsatile flow, and radius. The resulting microbubble dataset is then passed to the ultrasound simulation module, which uses the SIMUS linear ultrasound simulator \cite{garcia2024simus3}. Within this module, microbubble coordinates are segmented based on the transducer configuration and position. Microbubbles are treated as acoustic scatterers and organized into "scenes" according to user-defined parameters. For example, a mouse skull can be seeded with scatterers and overlaid with a microbubble scene to generate data that closely mimics \textit{in vivo} conditions. After ultrasound simulation, both radiofrequency (RF) and in-phase/quadrature (IQ) signals are retrieved, enabling downstream beamforming and image reconstruction.

%% Use \subsubsection, \paragraph, \subparagraph commands to 
%% start 3rd, 4th and 5th level sections.
%% Refer following link for more details.
%% https://en.wikibooks.org/wiki/LaTeX/Document_Structure#Sectioning_commands

\subsection{Microvascular Graph Geometries}
\label{sec:graph}

\begin{figure}[h]
    \centering
    \includegraphics[width=1.0\linewidth]{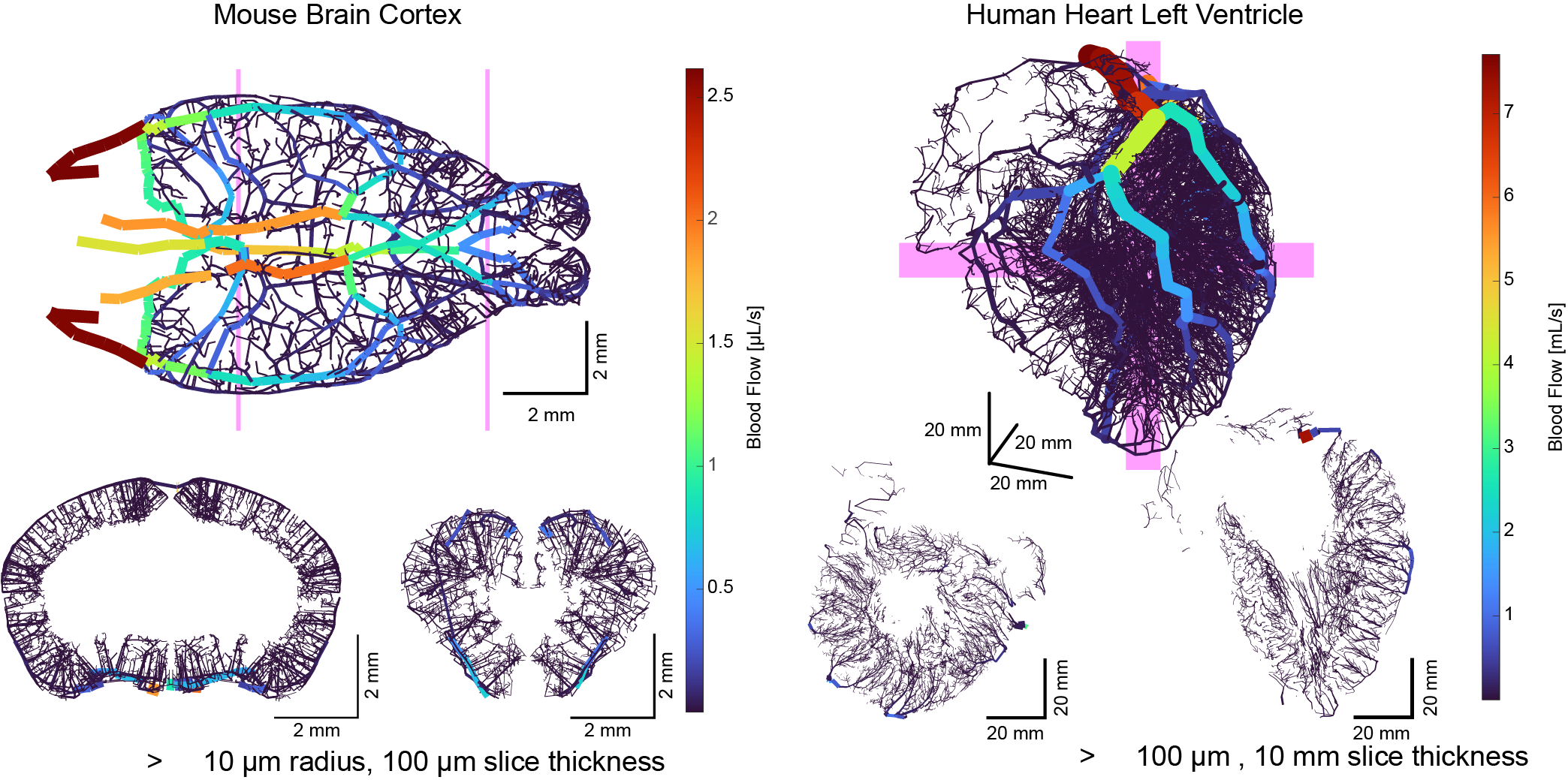}
    \caption{Mouse brain and human heart microvascular graphs with two sample cross-sections. Vessel radii greater than 10 \textmu m (slice thickness of 100 \textmu m) and 100 \textmu m (slice thickness of 10 mm) are displayed for the brain and heart, respectively.}
    \label{fig:2}
\end{figure}

In general, The-Bodega is designed to operate on directed graph networks composed of nodes and edges, enabling flexible and anatomically relevant simulation of microvascular flow. Any directed graph can be used as input to the simulator, allowing for broad applicability and data augmentation through randomized graph structures---a feature particularly useful for training AI models. The-Bodega currently includes four default vascular geometries: a half mouse brain model \cite{linninger2019mathematical}, a whole mouse brain, a human heart model \cite{schwarz2020topologic}, and a synthetic capillary network (used for checksum validation). In the case of the mouse brain and human heart, flow velocities were validated against \textit{in vivo} measured values, serving as an adequate backbone for simulating microbubble propagation. Coronal slice examples of the whole brain and human heart vascular networks are shown in Figure~\ref{fig:2}. 

The data for these graphs are publicly available in a dedicated repository (
%\url{https://ccdb.computecanada.ca/provostultrasoundlab/The-Bodega/DATASETS_public}
\url{to-be-published}) and consist of five files: flow data (.flo), node and edge structure (.nodes), spatial coordinates of nodes (.pos), optional pulse pressure values (.pul), and vessel radii at each edge (.rad). These graphs are saved snapshots of \cite{linninger2019mathematical} and \cite{schwarz2020topologic}. The directed graph architecture enables easy identification of input and output nodes and reduces redundancy in pathfinding by using the \texttt{allpaths} function, which guarantees uniqueness between microbubble tracks $T$. In the case of the mouse brain, this structure makes it possible to simulate microbubbles traversing every capillary within the network.

\subsection{Simulating individual microbubbles}
\label{sec:microbubbles}

After characterizing all possible travel paths, a microbubble distribution is simulated. In this work, we assume a Definity microbubble size distribution with a mean diameter of $\mu = 2~\mu$m and a standard deviation of $\sigma = 3~\mu$m. However, the framework supports both polydisperse and monodisperse microbubble distributions. Currently, the simulation considers microbubble diameter in relation to vessel size to determine flow behavior: if a microbubble's diameter exceeds the vessel diameter, it decelerates until it either shrinks sufficiently to pass through or disappears. A probabilistic weighting based on inlet node flow rate is applied to each trajectory, biasing the Monte Carlo simulation to favor paths with higher inflow.

Beyond steady-state flow conditions, we also precompute pulsatile flow waveforms based on user-defined parameters including heart rate, pulse wave velocity, and vessel compliance. Pulsatile waveforms are generated using the Pulse Decomposition Analysis (PDA) method \cite{baruch2011pulse}, in which five Gaussian functions are convolved with triangular windows to model the characteristic shape of an arterial pulse. However, we adapted the PDA model for propagation through a graph network and waveform accumulation to achieve steady-state values. This allows tuning of pulse morphology based on vessel-specific pulse pressure and radius.

The amplitude of each waveform is scaled by user-specified compliance values, applying the relation:

\begin{equation}
    \Delta V = C \cdot \Delta P
\end{equation}
where $\Delta V$ represents stroke volume (mL), $C$ is vascular compliance (mL/mmHg), and $\Delta P$ is pulse pressure (mmHg). Once waveforms are computed for each vascular edge, they are propagated in time using sub-pixel shifts. These shifts are determined by the length of trajectory $T$, the simulation time step $dt$, and the periodicity of the cardiac cycle. This results in a temporally accurate pulsatile velocity profile across the entire network, sampled over one or more cardiac cycles.

\begin{algorithm}
	\caption{Pulse decomposition analysis}
	\begin{algorithmic}[1]
        \State \textbf{Input:} Vascular graph $G$, pulse parameters, trajectory data
        \State \textbf{Output:} Edge-wise pulsatile waveforms
        \State Precompute edge shift ranges according to cardiac cycle length
        \State Precompute convolution pulse shapes
        \For{each edge $i$}
            \State Initialize empty waveform
            \For{each PDA component (5 total)}
                \State Retrieve or compute convolved waveform segment
                \State Add to composite waveform at appropriate delay
            \EndFor
            \State Normalize amplitude based on compliance
            \For{each range of shifts in edge $i$}
                \State Interpolate and sum the shifted waveform
            \EndFor
            \State Store average shifted waveform for edge $i$
        \EndFor
        \State Assign computed waveforms to graph edges
	\end{algorithmic} 
\end{algorithm}

\begin{figure}[h]
    \centering
    \includegraphics[width=0.7\linewidth]{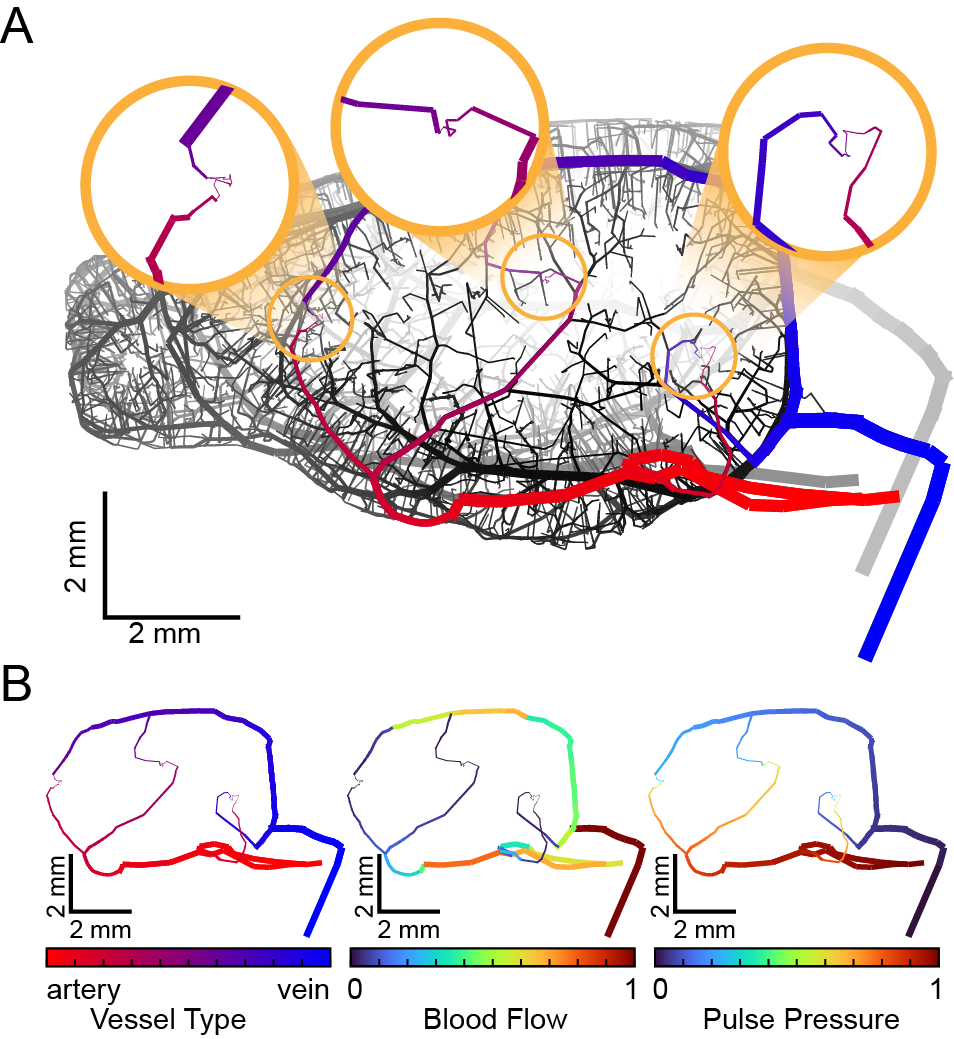}
    \caption{Microbubble populations are simulated through precomputed vascular paths. (A) Three representative nodal trajectories through the whole mouse brain network demonstrating the transition between artery, capillary, and vein. (B) Isolated 3~trajectories in panel A coded by vessel type, normalized blood flow, and normalized pulse pressure for computation of microbubble forward progression.}
    \label{fig:3}
\end{figure}

Finally, after precomputation of the possible trajectories, microbubble distribution, the pulsatile steady-state waveforms, and the possible Poiseuille flow approximations, a microbubble is then forward-propagated through the network. A trajectory, a microbubble, and a seed point are randomly chosen, with the seed point chosen preferentially at the beginning of the trajectory based on the probability density function over the trajectory radii (Figure \ref{fig:3}A). The validity of configuration is screened based on the flow and transit time across the network. The microbubble, based on velocity, pulsatile flow, and radii, is propagated forward as in \cite{belgharbi2023anatomically} using precomputed steady-state values for each specific track (Figure \ref{fig:3}B). Here, three full capillary transit tracks are shown originating from two arteries at the Circle of Willis and terminating at the same vein. As such, we have ground truth information on the vessel architecture, the underlying blood flow, and the pulsatile pressure of the PDA waveform.

Afterwards, the microbubble path along the centerline is altered so that a smooth trajectory runs along it, scaled by the radius. To do so, the centerline is first Savitzky--Golay filtered (5~window, 2nd order) and the Rotation Minimizing Frames (RMF) algorithm was implemented to avoid sudden frame of reference shifts along the trajectory \cite{10.1145/1330511.1330513}. The RMF algorithm avoids discontinuity by minimizing the perpendicular planes distributed along a space curve via relative rotation and the double reflection method, a computationally efficient algorithm using two reflections to approximate the RMF. From here, a random radial distance is applied and the trajectory is normalized by $1/\sqrt{1-p}$, where $p$ is the Poiseuille factor, and multiplied by the radius along the trajectory $T$. Finally, if a microbubble encounters a vessel path that is too small, it will disappear. As a result, thousands of individual microbubbles can be simulated for further combination into a specified dataset. The-Bodega features both serial and parallel CPU processing as well as batched simulations for adaptability to high-performance computing clusters (HPCs).

\subsection{3D Microbubble Dataset Formation}
\label{sec:microbubble_dataset}
Once all microbubbles are simulated and stored in a temporary scratch directory, they are combined into datasets that emulate \textit{in vivo} imaging experiments. Each dataset row represents a ``line'' of multiple microbubbles, temporally aligned based on the cardiac cycle and spanning durations of up to several minutes. More formally, let the full set of simulated microbubbles be:
\[
\mathcal{B} = \{b_1, b_2, \ldots, b_{N}\}
\]
Then, for each of $M$ frames $f$, we define a random subset:
\[
\mathcal{B}_f \subset \mathcal{B}, \quad |\mathcal{B}_f| = M
\]
such that:
\[
\mathcal{B}_f \sim \text{Sample}(\mathcal{B}, M; \text{without replacement})
\]
These lines are written to disk using low-level HDF5 functions, providing greater flexibility for use across different programming environments. The code supports both parallel and batched processing on HPCs.

\begin{figure}[h]
    \centering
    \includegraphics[width=0.8\linewidth]{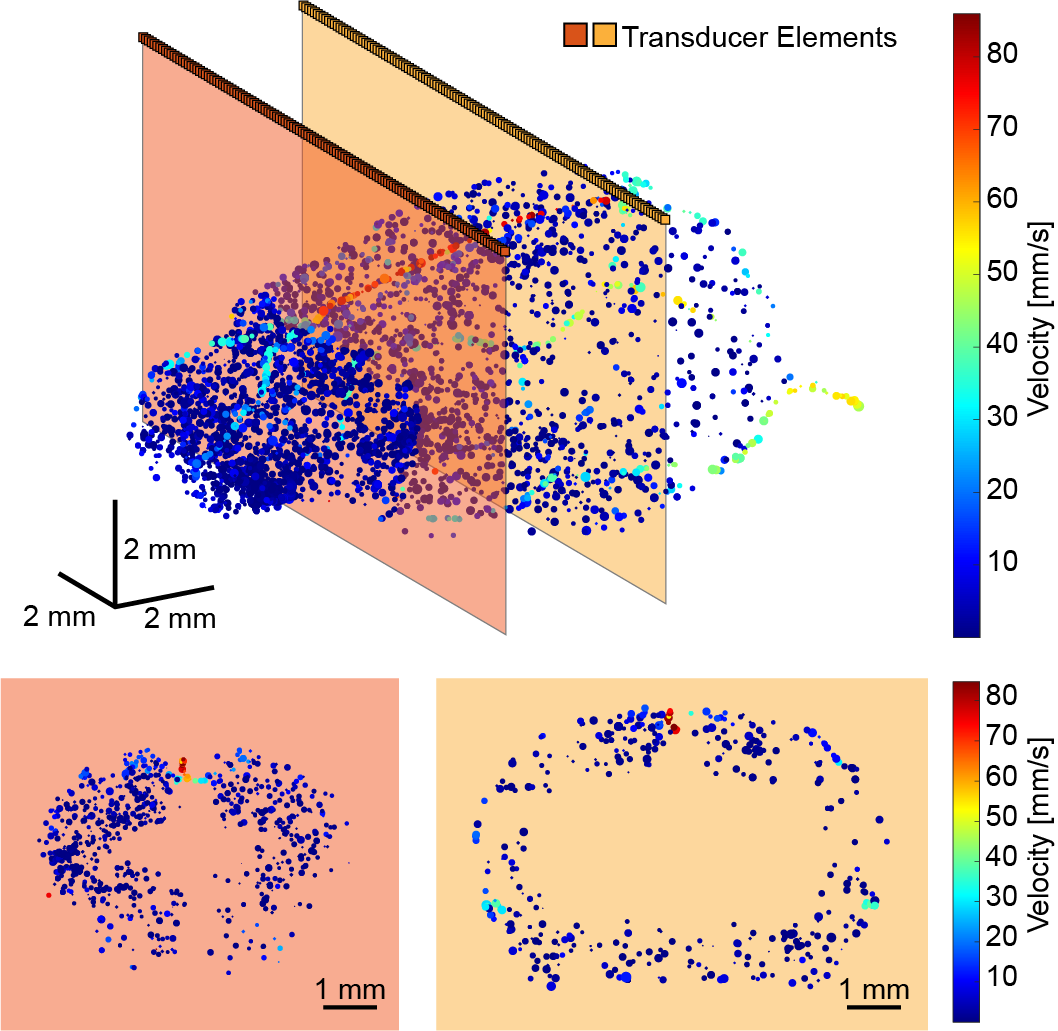}
    \caption{Resultant combination of microbubbles populated for one mouse whole brain dataset. Two slices, dictated by the field of view of a linear array, are chosen to display cross-sectional views of the microbubbles in 2D.}
    \label{fig:4}
\end{figure}

To balance efficiency and data readability, we chose to use the HDF5 file format for its high-performance flexible structure. This structure enables dynamic control over the number of bubbles per line and supports strided access, making it suitable for memory-constrained environments where loading the entire dataset at once is impractical. Additionally, chunk size and compression level can be changed to either ensure efficient data access or decrease the resultant file size. Example trajectories of flowing microbubbles are shown in Figure \ref{fig:4} and Video S1. Here, two slices dictated by the position of the ultrasound transducer (described in Section \ref{sec:ultrasound_sim}) show the size and velocity distribution of microbubbles in the whole brain cortex.

\subsection{Configuring the Ultrasound Simulation Module}
\label{sec:ultrasound_sim}

To simulate ultrasound data, we chose to implement the flexible, linear acoustic simulator SIMUS, based on previous work by Garcia et al. \cite{garcia2024simus3}, but the class instance is constructed to mimic the format provided by the Verasonics Vantage and NXT SDK. Moreover, SIMUS is conducted in the frequency domain and is particle-based such that the simulation does not depend on the Cartesian grid for simulations. All simulations were performed in 3D, even for 2D data, to mimic in- and out-of-plane microbubble movement using GPU-accelerated CUDA frameworks (NVIDIA RTX 3090, 24~GB vRAM). 

The module begins by defining the transducer used for acquisition that we would like to simulate, with initial parameters contained in JSON format. This includes parameters such as medium attenuation, number of scatterers per cell, and the microjitter amplitude. The-Bodega contains transducer configurations by default for low-frequency (2--4~MHz) and high-frequency probes (15~MHz) 2D probes, as well as a high-frequency row-column array (RCA) \cite{flesch20174d,morton2003theoretical}. Subsequently, the transducer becomes the origin of the ultrasound simulation; thus, the slice of the microbubble dataset is dictated by the position of the transducer.

To simulate \textit{in silico} RF and IQ data, we compose different "blocks" seeded with scatterer points. For example, to compose a ULM acquisition for transcranial \textit{in vivo} mice, we can take 3D micro-CT scans from an open-source dataset \cite{FACEBASE:V92} to seed scatterer points where their reflection coefficient and the density positions are determined by a random distribution of the cumulative probability of the intensity of the micro-CT. A second block can be composed for the microbubbles flowing through the field of view (FOV) of the chosen transducer and co-aligned with the previous tissue and skull clutter block. Each block is then simulated serially or in parallel, compatible with compute clustering, and individual buffers of RF or IQ data for each block can be summed to generate the full ultrasound dataset. 

To draw parallels, in the human heart, this process can be used to segment different views of the heart (coronal, 4-chamber, etc.) and tissue clutter can be added in the same manner by seeding scatterers in accordance with the intensity of the B-mode \cite{garcia2024simus3}. The separation between the microbubbles and the tissue clutter allows for comparison of spatiotemporal filtering methods with clean signal. Finally, any user-defined beamformer can be used for image formation. For the duration of this paper, images are shown using conventional GPU-accelerated delay-and-sum (DAS).

\subsection{ULM processing}
\label{sec:ulm}
Ultrasound localization microscopy was performed using the TAL algorithm \cite{leconte2024tracking} with Kalman filtering and radial symmetry. Briefly, tracks were constructed through vesselness filtering in the spacetime dimension and centerline-thinned to create \textit{a priori} tracks that were then refined through radial symmetry and Kalman filtering. Track density maps were created through the accumulation of multiple buffers of simulated ultrasound data. Velocity maps were constructed by taking the mean projected value over all buffers. 

For ULM tracking with the RCA, images were constructed using orthogonal plane wave compounding and delay and sum beamforming. IQ volumes were correlated with the PSF simulated using the SIMUS simulator. MB signals were enhanced through lag-1 autocorrelation and temporal ensemble averaging with a hanning window of 9 frames. To detect the local maxima, subpixel localization was performed through 3D gaussian fitting with a kernel of size 9. Here, local maxima with correlation coefficients lower than 0.4 were rejected. To obtain super-resolved trajectories, we used the Hungarian method \cite{kuhn2004hungarian} with a maximum linking distance of three voxels (0.1925~mm) between two consecutive frames corresponding to a maximal velocity around 100~mm/s and without gap filling. 

Dynamic ULM (dULM) for visualizing pulsatility in 3D was performed through temporal realignment over slow time \cite{wu11124235,ghigo2024dynamic}. Here, in lieu of ECG signal, the tissue Doppler used to spatiotemporally align microbubble phase in the cardiac cycle during the simulation was used for realignment. Here, each ultrasound dataset was split into two 2 cardiac cycle segments in which both segments were realigned to a reference frame. Realigned tracks were then smoothed using cubic splines and velocities were obtained from the closed-form derivative of the spline curve. This resulted in a temporally-varying dULM cineloop where the traveling pulse wave can be visualized.

Functional ULM (fULM) was performed using the methodology developed by \cite{renaudin2022functional}, where the activation maps were constructed by pixel-by-pixel Pearson correlation analysis between the ULM density 2D matrices in slow time and the generated pulse train. A sliding window of 4~s was used with a stride of 0.4~s to generate cineloops of ULM density. These 3D matrices were then temporally realigned to average signal from 5~pulses to a single pulse stimulation. Correlation maps were then overlaid onto the density maps to create fULM images.

\subsection{Capillary dynamics and ablation analysis}

Due to the number of edges and nodes contained in the mouse brain vascular graph, the analysis of capillary reconstruction via ULM was performed using a 2~mm $\times$ 2~mm section of the graph. Saturation curves (i.e., the overlap between reconstructed images and the ground truth network) were generated to illustrate how rapidly a ULM acquisition can populate pial and penetrating arterioles, venules, and the capillary mesh network as a function of time and the concentration (between 378 to 1511 $MB/mm^3$) of circulating microbubbles. Since it takes minutes to fully reconstruct the capillary network, we ranked the capillary nodes by importance to estimate the number of capillary tracks needed to functionally sample the capillary mesh. This importance was composed of a linear combination of the betweenness centrality and PageRank estimator for node connectivity. Node score saturation curves were generated by accumulating the total score or overlap of the reconstructed ULM image with important capillary nodes.

Skull clutter was simulated using 11~scatterers per cell, injected with 5~$\mu$m of jitter movement. The resultant ultrasound data consisted of a linear combination of the skull clutter and the microbubble signals. These images were spatiotemporally filtered using a singular value decomposition (SVD) clutter filter, and the Dice score, Jaccard index, sensitivity, and specificity were calculated on binarized images between reconstructed SVD-filtered data and microbubble-only data after ULM tracking. Two conditions were tested with different eigenvalue-to-ensemble size cutoffs: [5/500] and [25/500]. 

Thus, the conditions tested are as follows: ground truth (taken directly from the input graph), microbubble-only (MB-only) where vessels are constructed after ultrasound simulation and ULM tracking, and SVD-filtered after ultrasound simulation but before ULM tracking. Saturation curves were calculated based on the microbubble-only positional data compared to the ground truth.

\subsection{Simulating Neurovascular responses}

\begin{figure}[!h]
    \centering
    \includegraphics[width=0.7\linewidth]{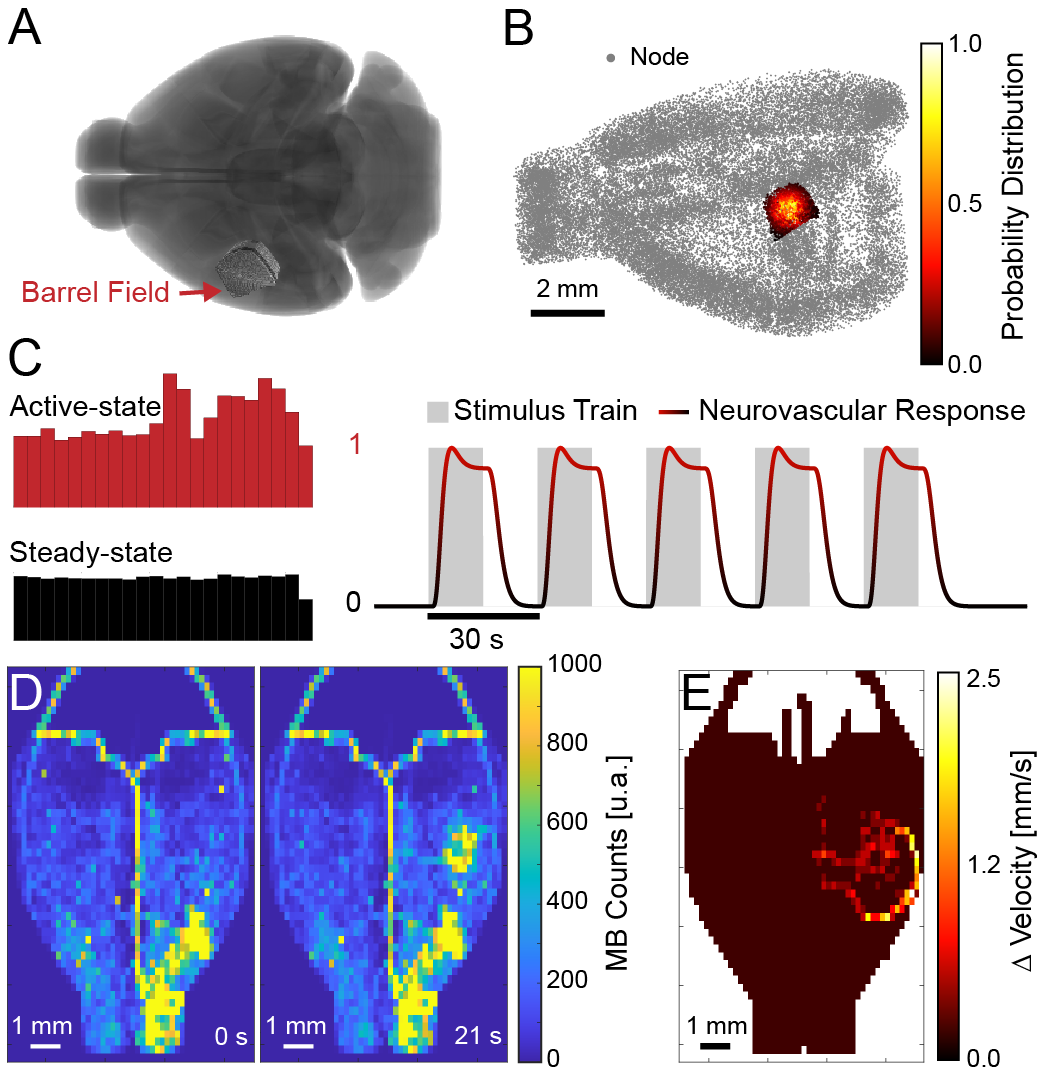}
    \caption{Simulating microbubbles flowing through neurovascular-activated barrel field cortex. (A) Allen Brain Atlas common coordinate framework reconstruction with barrel field cortex masks. (B) Augmented probability distribution of nodes found within the barrel field region of interest (ROI) in panel A. (C) Two microbubble distributions simulated in steady-state and active-state environments. Resultant dynamic dataset is resampled according to the simulated convolved hemodynamic response. (D) Maximum intensity projections of microbubble count density from the top view of the mouse whole brain cortex in steady state (left) and active state (right). Images shown are summed in the temporal direction over 5~stimulations. (E) Changes in velocity at full activation compared to baseline steady state, indicating increased cerebral blood flow (CBF).}    
    \label{fig:5}
\end{figure}
Hemodynamics play a large role in neuronal homeostasis, and current neuroimaging techniques, such as fMRI, functional ultrasound, and functional ULM, rely on the neurovascular coupling phenomenon for analysis of evoked and resting-state vascular responses. Thus, we chose to develop a simple evoked-hemodynamic model for benchmarking fULM techniques. Here, we detail the methodology for simulating cerebral blood volume (CBV) and cerebral blood flow (CBF) augmentations in the barrel field (BF); however, any brain region or connected brain regions can be stimulated with this framework.

Using the annotated Allen Brain Atlas Common Coordinate Framework \cite{wang2020allen}, we were able to extract binary volumes containing parcellated hemispherical BF regions (Figure \ref{fig:5}A). This barrel field cortex is composed of layers 1--6. To simulate right whisker stimulation, we chose to apply neurovascular changes to the left BF cortex with a spatial distribution of CBF/CBV changes as a Gaussian function from the ROI center (Figure \ref{fig:5}B). The nodes within this ROI volume are then segmented into capillary, pre-capillary arterioles, and penetrating arterioles, and separate flow and radial segments are augmented with physiological values from the literature \cite{hillman2014coupling,srinivasan2011optical}. Additionally, the trajectory selection probability distribution was augmented to preferentially increase vascular paths that lead to the BF cortex to simulate increased CBV.

To dynamically simulate functional hemodynamic responses to stimulus-evoked impulses, two microbubble distributions were simulated: steady-state and full whisker activation (Figure \ref{fig:5}C, left). Figure \ref{fig:5}C (right) shows a hemodynamic response function simulated based on an input pulse train to determine distribution sampling from state 0 (steady-state) to state 1 (whisker-stimulated). The hemodynamic response function was modeled by defining a stimulus square wave signal (15~s off, 15~s on, repeated 5~times) and modeling the neural and neurovascular coupling response described in \cite{BUXTON2004S220}. Briefly, the neural response was modeled as a set of two equations that describe a simple inhibitory feedback system as a function of the excitatory and inhibitory input. This neural response trace was then convolved with the hemodynamic impulse response modeled as a gamma-variate function with full width at half maximum of 4~seconds and an approximate delay of 1~second. This neurovascular response curve is then treated as a cumulative probability density function that varies in time to determine from which microbubble distribution is selected to create the full dataset as described in Section \ref{sec:microbubble_dataset}.

Thus, both CBV (Figure \ref{fig:5}D) and CBF (Figure \ref{fig:5}E) increases can be demonstrated simply by augmenting the input microvascular graph to generate ground truth data for contrast-enhanced ultrasound imaging. Note that our model omits flow redistribution in the altered state; thus, flow balancing can be performed through the process detailed in \cite{linninger2019mathematical} to redistribute the altered flow network, leading to increased biological realism.

\subsection{Applying Motion to the Heart}

\begin{figure}[h]
    \centering
    \includegraphics[width=\linewidth]{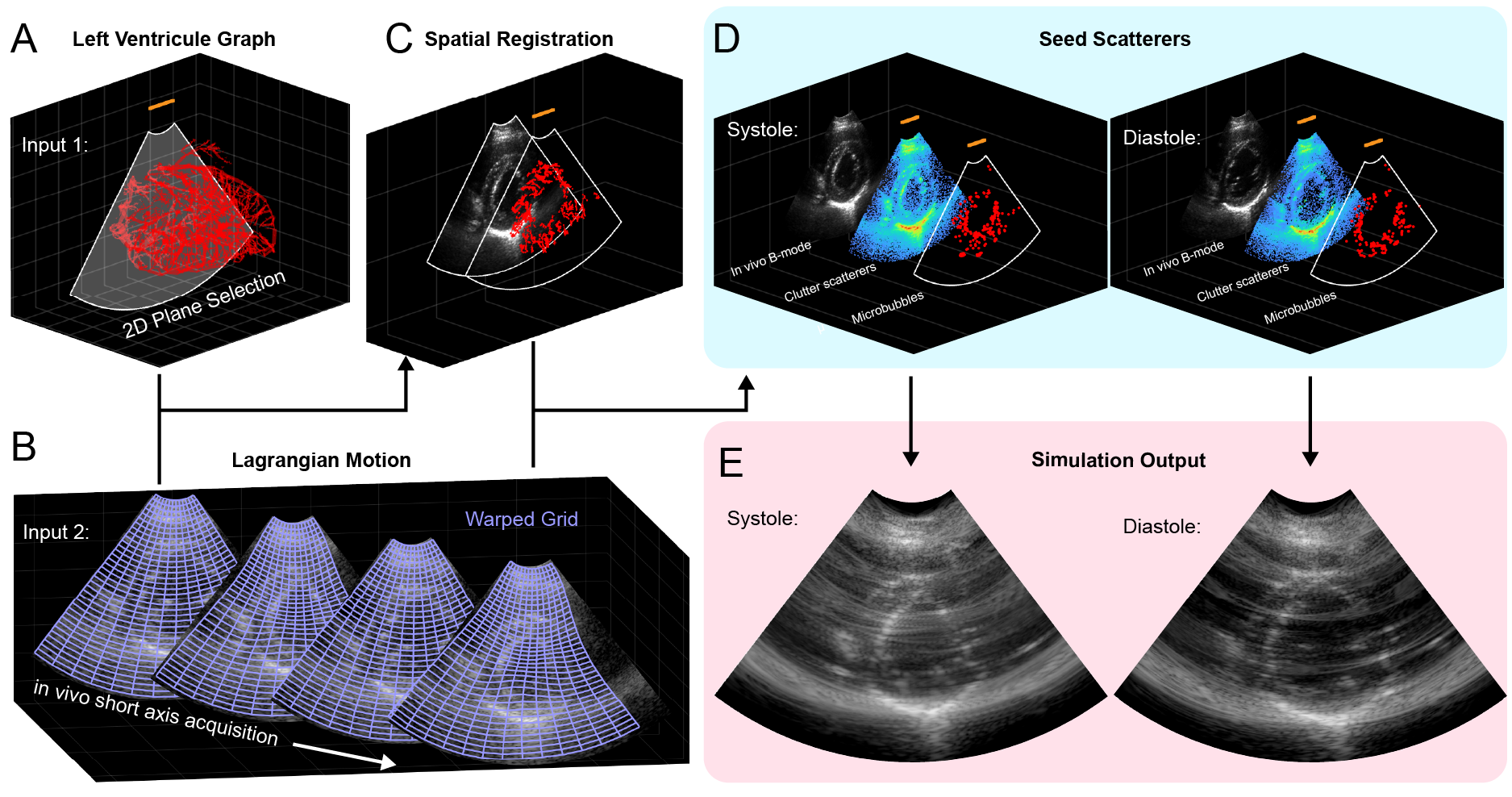}
    \caption{Schematic representation of the simulation pipeline designed to integrate dynamic motion into static cardiac simulations. (A) The first input features a 3D reconstruction of the human heart's microvascular left ventricle graph, with a selected 2D imaging plane. (B) The second input shows an \textit{in vivo} B-mode short-axis acquisition, accompanied by the associated Lagrangian motion grid. (C) Spatial registration of the selected 2D imaging plane with a specific frame of the in vivo B-mode short axis. (D) Illustration of seeded scatterers with backscatter intensity proportional to the B-mode, and seeded microbubbles within the 2D slice of the graph where the motion field is applied, representing both systole and diastole phases. (E) Simulation output depicting B-mode images for systole and diastole phases.}
    \label{fig:cardiac_methods}
\end{figure}

Inherent motion present in all biological tissues and organs creates a significant obstacle in ultrasound imaging. ULM, in particular, is highly dependent on how effectively tissue signals are separated from microbubble signals. The SVD is the most widely adopted algorithm currently for clutter filtering. However, it relies on the assumption that tissue exhibits lower motion than microbubbles, leading to higher spatiotemporal coherence in the tissue signal. Since clutter filtering approaches using the SVD operate under the assumption that spatiotemporal decorrelation can be adequately represented within finite eigenspaces, This assumption rapidly fails in moving organs such as the heart, thereby compounding the challenges of performing ULM in a \textit{in vivo} cardiac environment.

To develop a framework for exploring ULM in the human heart under motion and realistic clutter, we first selected a short-axis plane from the 3D human left ventricle microvascular graph (Figure~\ref{fig:cardiac_methods}A, Input 1). Subsequently, we captured an \textit{in vivo} human heart acquisition with a short-axis view, consisting of both diastolic and systolic motions, as input 2 in Figure~\ref{fig:cardiac_methods}B. The corresponding diastole and systole motion fields were then computed using the methodology detailed in \cite{cormier_heart}. Briefly, a Lagrangian coordinate system was derived by estimating tissue motion from Doppler velocity vector measurements through an iterative process until convergence. The selected short-axis plane from Figure~\ref{fig:cardiac_methods}A was then spatially registered to a specific frame of the \textit{in vivo} dataset (Figure~\ref{fig:cardiac_methods}C). This process provides a temporally varying motion grid coherent with the aligned \textit{in vivo} echocardiography. This motion can then be applied to the microbubbles flowing within that cross-sectional slice, displacing them in synchronization with the cardiac motion. Consequently, the microbubbles are displaced in coordination with the tissue clutter scatterers to simulate realistic systolic and diastolic cycle phases (Figure~\ref{fig:cardiac_methods}D).

Ultrasound data can be simulated as described in Section~\ref{sec:ultrasound_sim}, where representative corresponding beamformed images are shown in Figure~\ref{fig:cardiac_methods}E for diastole and systole. To investigate the effects of SVD in the presence of tissue clutter and motion, we isolated microbubble signals using an SVD spatiotemporal clutter filter. ULM was then performed using the TAL framework \cite{leconte2024tracking} to reconstruct a representation of the microvasculature. Due to the presence of motion, a static image was acquired by inverting the motion previously applied to the scatterers, effectively realigning the microbubble positions after ULM tracking estimation to temporally align all ULM trajectories to a common reference frame. Thus, the contribution of motion and tissue clutter to tracking and localization errors can be independently analyzed.

\section{Results}

\subsection{Computational Time Benchmarking}

Dynamic microbubble simulations were tested on a Linux distribution (Ubuntu 24.04 LTS) with 8~CPU cores, 256~GB RAM, and a 24~GB GPU (NVIDIA RTX 3090). In general, the simulation can be performed serially, but performance improvements using combinations of CPU and GPU parallel processing can rapidly accelerate The-Bodega. Additionally, since we perform sequential Monte Carlo simulations that may exceed RAM constraints, we save individual microbubble simulated results and individual lines of microbubbles for dataset generation as individual HDF5 files. Thus, disk speed can have a large effect on simulation runtime. Here, we use a 4~TB PCIe 3.0 NVMe SSD for all computational storage.

\begin{table}[h]
    \centering
    \begin{tabular}{p{.19\linewidth}|p{.19\linewidth}|p{.16\linewidth}|p{.18\linewidth}|p{.09\linewidth}}
         \textbf{Benchmark} & \textbf{Whole Brain Cortex} & \textbf{Half Brain Cortex} & \textbf{Human Heart} & \textbf{Toy} \\
         \hline
         Memory (GB) & 55.153 & 10.573 & 4.572 & 4.303 \\
         Total Time (s) & 2844.85 & 3023.69 & 2090.09 & 797.33
    \end{tabular}
    \caption{Memory usage (GB) and total script runtime across domains.}
    \label{tab:memory_usage}
\end{table}

The total computational time and memory usage are summarized in Table \ref{tab:memory_usage} for each separate vascular network and example script available in the pipeline. Here, 32,000~microbubbles were simulated at 10~kHz sampling rate for a total of 180~seconds of microbubble flowing data. From this, microbubbles were randomly sampled to create volumes of 5000~microbubbles per frame for a total of 180~s of data. As expected, the total required memory scales with vascular graph complexity, requiring a maximum memory of 55~GB to simulate the whole brain cortex (approximately 2.1~million nodes and 2.8~million edges). We observe that the total computation time does not scale in the same manner, with the half brain and whole brain cortex taking approximately the same amount of time despite the twofold difference in nodes and number of trajectories.

\begin{figure}[h]
    \centering
    \includegraphics[width=0.79\linewidth]{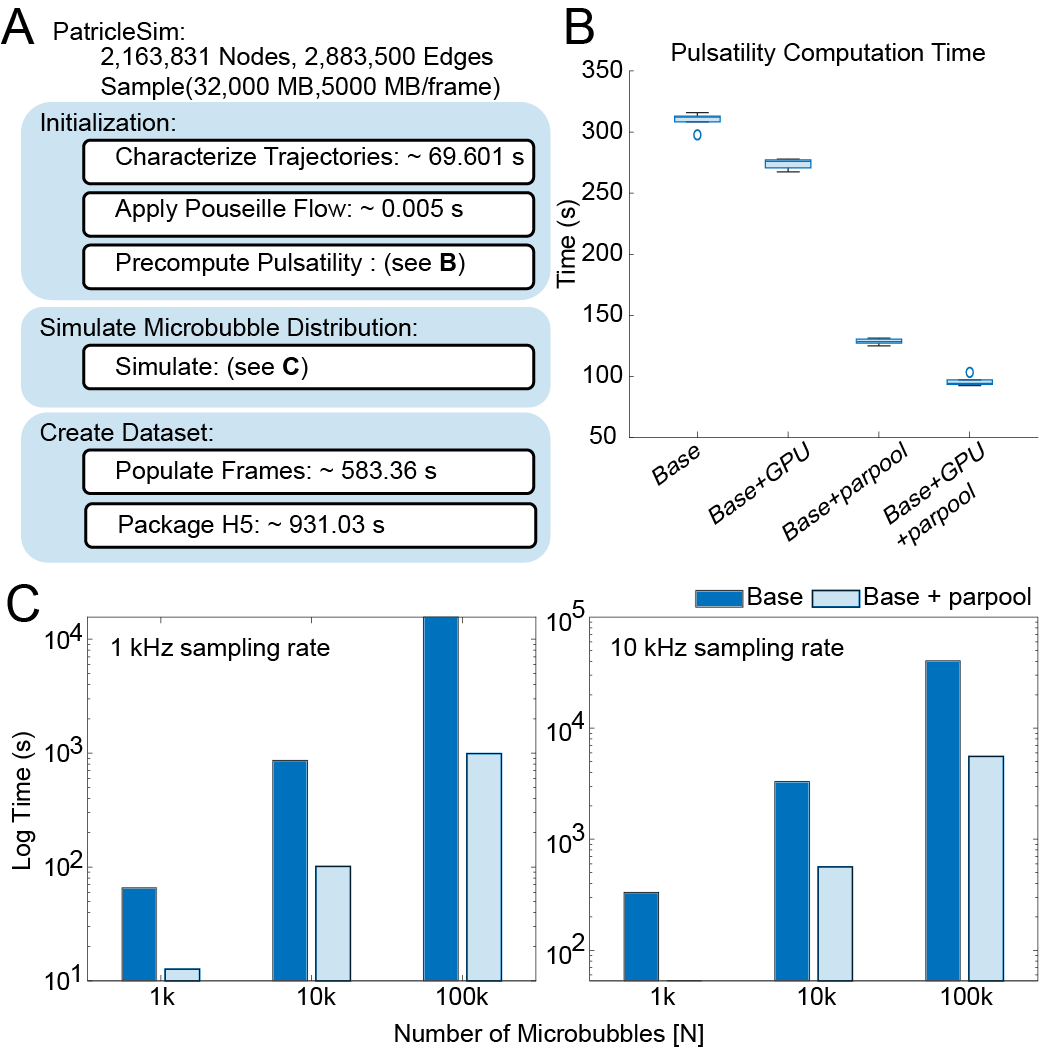}
    \caption{Computational time for the ParticleSim class and the mouse whole brain cortex graph model. (A) Flow diagram and corresponding time estimates for each component of the particle simulator. (B) Steady-state pulsatility computational time comparison between serial and parallel processing configurations. (C) Logarithmic computational time for 1~kHz and 10~kHz sampling rates as a function of the number of requested microbubbles.}
    \label{fig:6}
\end{figure}

Figure \ref{fig:6}A illustrates in more detail the computational complexity and computational time of the particle simulator module. By far, the greatest computational burden in the initialization stage is the characterization of the trajectories, or rather, the retrieval of all possible pathways in the vascular graph (MATLAB's \texttt{allpaths} function). Thus, for networks that are fully connected, such as the brain graphs, the number of pathways far outnumbers other available vascular structures. Input of an undirected graph in these cases will be unretrievable due to the numerous loops. Therefore, it is recommended that input vascular graphs be directed, limiting the number of possible pathways. 

The second computational burden is the calculation of steady-state pulsatility using the PDA algorithm. Since this is calculated for each edge in the graph, previous implementations required upwards of 10~minutes for calculation. Here, vectorization operations for edge retrieval and approximation of the sample shift along the trajectory pathway decreased computational time by 76\%. Additional speedup was achieved through the implementation of GPU arrays for faster sorting over millions of edges as well as an edge-wise parallel operation (threads parpool environment) for the Gaussian summation and subsample shifts according to the input cardiac cycle (Figure \ref{fig:6}B). 

Finally, we demonstrate the computation time for two sampling rates, 1~kHz and 10~kHz. Figure \ref{fig:6}C illustrates the time needed to simulate using the base model or with the parallel configuration under the conditions in Figure \ref{fig:6}A as a function of the number of microbubbles simulated in the distribution. For the 1~kHz case, it takes 1.08~min, 14.40~min, and 260.15~min for serial simulation of 1K, 10K, and 100K microbubbles, respectively. In contrast, it takes between 0.2 and 16~minutes to simulate the same microbubble range using CPU parallel processing. Simulating the same conditions at 10~kHz increases the computational time linearly by a factor of 3--5.

\subsection{3D Pulsatility in dynamic ULM via Row-Column Arrays}
\begin{figure}[h]
    \centering
    \includegraphics[width=\linewidth]{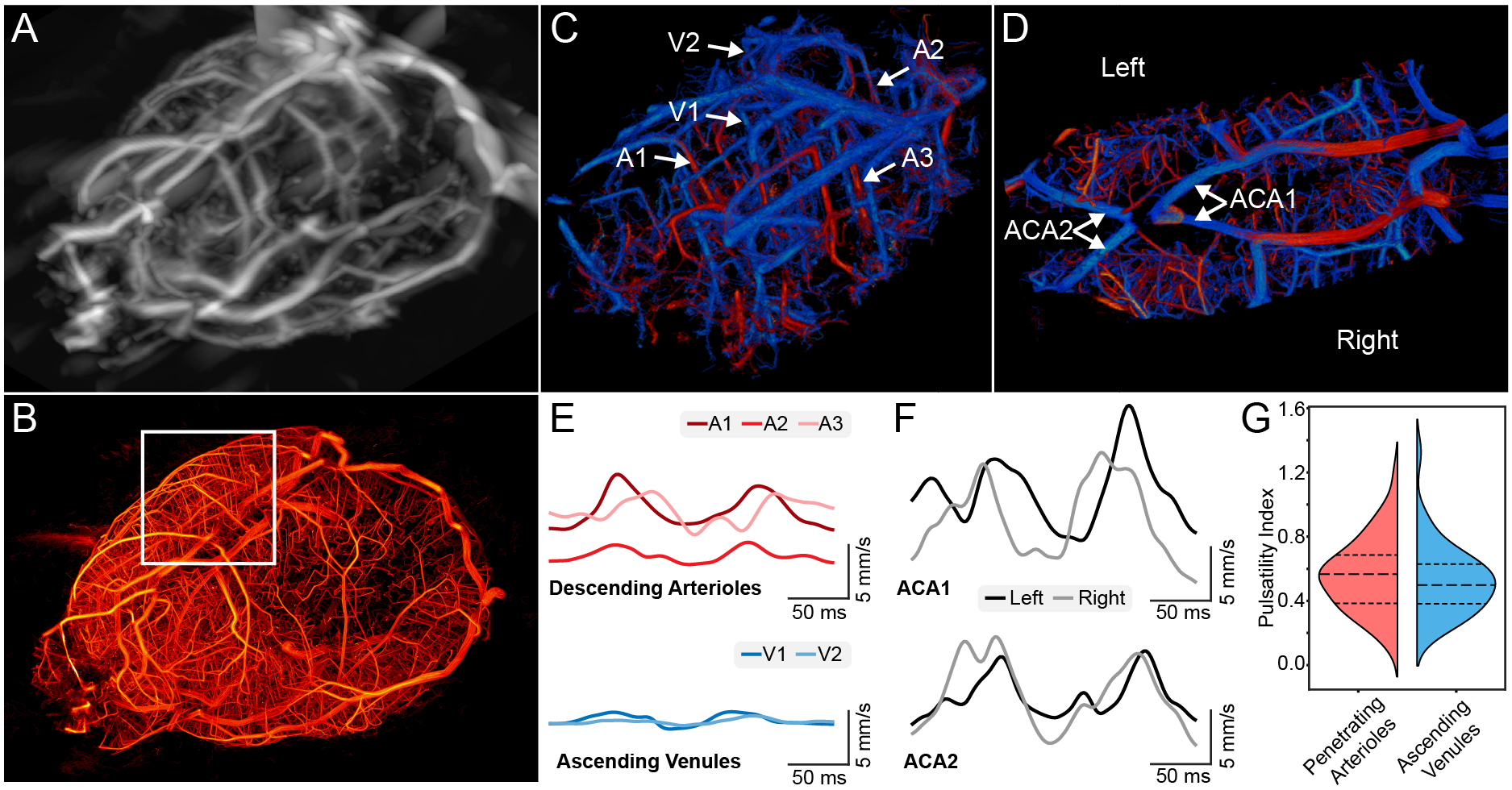}
    \caption{3D pulsatile flow in dynamic ULM via Row-Column Arrays in the Mouse Whole Brain Cortex. (A) CEUS reconstruction of the mouse brain. (B) ULM reconstruction after localization and tracking. (C) Sample vascular segment in B with signed directional flows (red - down, blue -up). Penetrating arterioles are labeled with A and ascending venules are labeled with V. (D) bottom view of the Circle of Willis with labeled anterior cerebral arteries. (E) Dynamic velocity traces for arterioles and venules labeled in C. (F) Sample velocity measurements at the ACA in D. (G) Violin plot of pulsatile index for arterioes (n = 52) and venules (n = 41).}
    \label{fig:RCA_results}
\end{figure}

Pulsatility is another important facet of cerebral blood flow that may give indication towards- or contribute to the progression of neurological and cardiovascular diseases. By temporally processing static ULM images, dynamic ULM provides the ability to noninvasively assess the pulsatile flow in small vessels like descending arterioles and ascending venules. The differences of which may give indication to capillary fragility as well as overall brain health and function. As a demonstration, here we show how 3D dULM performed through a RCA can be used to assess the simulated pulsatile flow in our hemodynamic mouse brains.

The difference between CEUS and ULM with the RCA is illustrated in Figures \ref{fig:RCA_results}A and B. Here, In the contrast-enhanced Doppler volume (Figure \ref{fig:RCA_results}A), only the largest cerebral vessels are clearly visible, highlighting the spatial resolution limits of CEUS in this context. Thus, only assessing pulsatility in these vessels are possible in CEUS. In contrast, ULM imaging (Figure \ref{fig:RCA_results}B) provided a highly-detailed vascular map, demonstrating reconstruction of the larger Circle of Willis down to the smaller penetrating arterioles and ascending venules. 

Here, dULM was constructed (Supplemental Video S2) and the flow direction of the microbubbles can be used to determine vascular architecture in the cortex. The signed density map of the ROI (Figure \ref{fig:RCA_results}B) of the cortex (Figure \ref{fig:RCA_results}C) reveals the flow directionality, with ascending venules shown in blue and penetrating arterioles in red. Thus individual small vessels were automatically segmented to evaluate dynamic velocity changes over time. Pulsatile velocity waveforms were extracted after temporal realignment across two cardiac cycles for three arterioles and two venules (indicated by arrows in Figure \ref{fig:RCA_results}C). The same procedure was performed for the Circle of Willis at the bottom of the mouse brain (Figure \ref{fig:RCA_results}D) to evaluate the ACA at two different points and whether they supply the left or the right cerebral arteries.

Figure \ref{fig:RCA_results}E demonstrates individual velocity traces for three segmented arterioles and two segmented venules, as indicated in Figure \ref{fig:RCA_results}C. Here, we can identify two peaks in velocity for both arterioles and venules, indicative of two simulated cardiac cycles. However, as expected, we see a lower pulsatile change in venules as compared to arterioles. Conversely, at the ACA, we see higher pulsatile velocity changes (Figure \ref{fig:RCA_results}F). We are still able to identify the two cardiac cycles and we also observe a small shift in the phase of the velocity peaks at points ACA2 compared to the preceding points at ACA1. Lastly, we sample and calculate the pulsatility index (PI) for (n = 52) penetrating arterioles and (n = 41) ascending venules in the cortex across the whole brain. Indeed, we see that the mean PI is higher in  arterioles as opposed to venules. Importantly, this characterizes the use of the PDA algorithm for propagating the pulse wave through these microvascular network and validates a model of pulsatility that mimics \textit{in vivo} pulsatile flow. This further illustrates the utility of The-Bodega for dynamic microbubble simulated environments.

\subsection{Assessing capillary vessel dynamics and ULM reconstruction}

\begin{figure}[h]
    \centering
    \includegraphics[width=.9\linewidth]{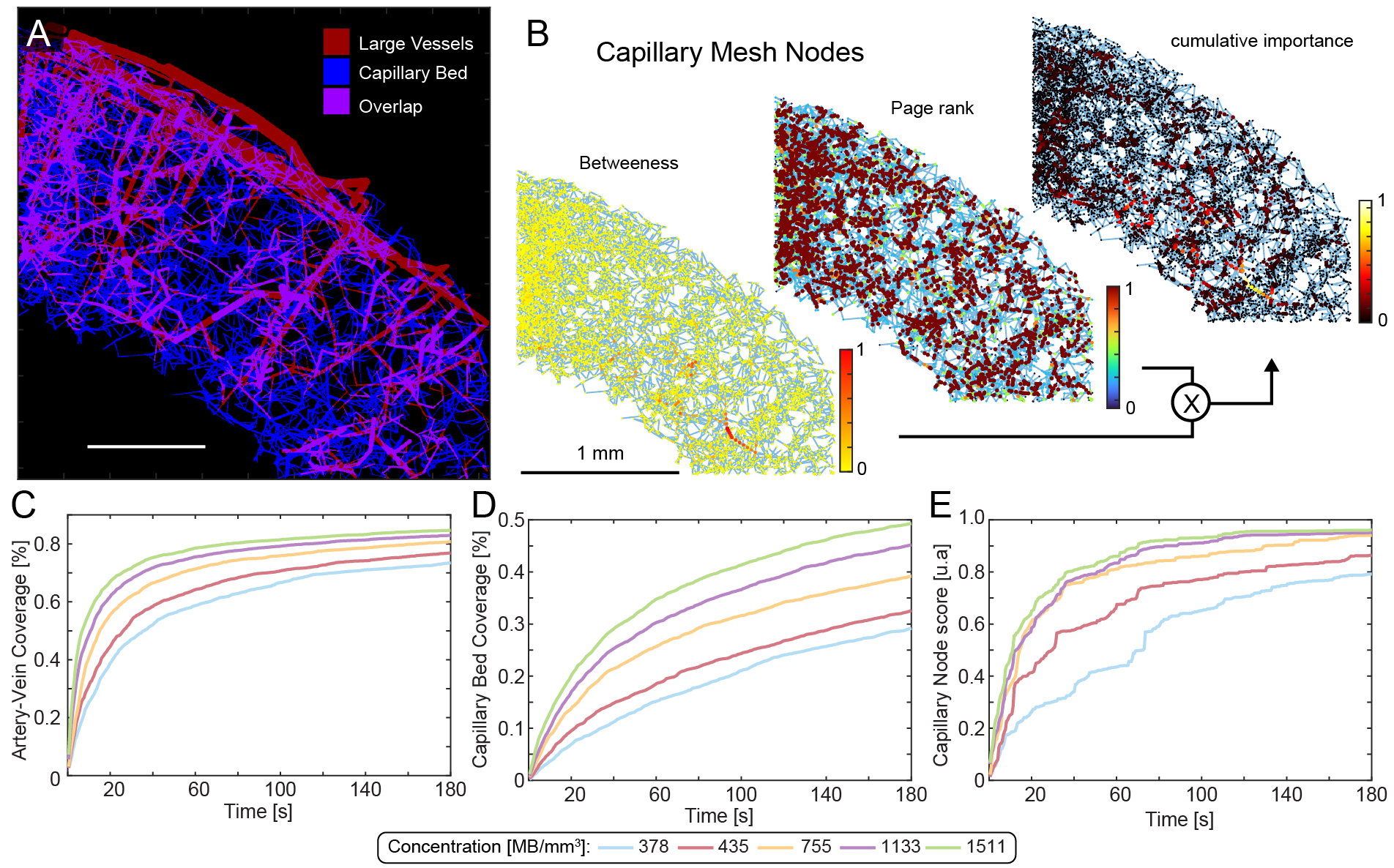}
    \caption{Capillary mesh ULM analysis. (A) Ground truth 2D projection of the vascular graph separated by large vessels and the capillary mesh. (B) Capillary mesh connectivity analysis using normalized betweenness centrality and PageRank importance measures. (C) ULM saturation curves as percentage of ground truth large vessels. (D) ULM saturation curves as percentage of the ground truth capillary mesh. (E) ULM score curves weighted by capillary node importance saturation.}
    \label{fig:7a}
\end{figure}

\begin{figure}[h]
    \centering
    \includegraphics[width=.9\linewidth]{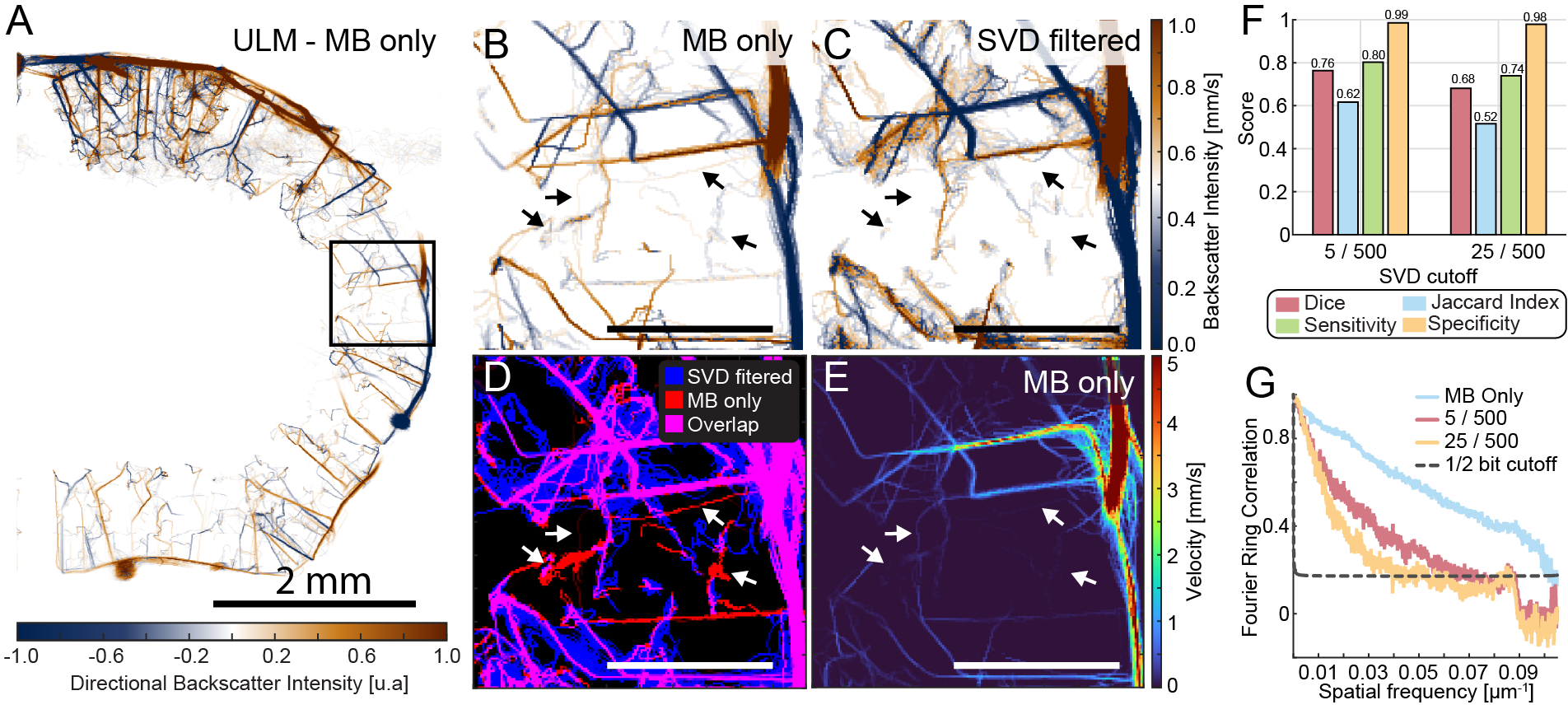}
    \caption{Capillary ablation via spatiotemporal SVD filtering. (A) Representative ULM maps of microbubble-only tracking over 180~s at 378~microbubbles per mm$^3$. (B) Zoomed image of microbubble-only tracking from F. (C) Zoomed image with added skull clutter and SVD clutter filter from F. (D) Overlay between microbubble-only zoomed image and SVD-filtered zoomed image. (E) Microbubble-only velocity map. Arrows in panels B--E indicate individual capillary tracks and the scale bar = 500 $\mu m$. (F) Dice score, Jaccard index, sensitivity, and specificity measurements for two different eigenvalue cutoff filter values. (G) Fourier ring correlation for microbubble-only images compared to two eigenvalue SVD cutoff reconstructions, with the half-bit threshold intersection indicating resolution.}
    \label{fig:7b}
\end{figure}

A pressing question in ULM is whether capillary perfusion within the brain can be accurately reconstructed, given the vast number of capillary vessels and the rarity of multiple microbubbles passing through the same capillary vessel within minutes \cite{hingot2021measuring}. To demonstrate the utility of large-scale microbubble simulations through the mouse brain, we performed two experiments examining vessel saturation (Figure \ref{fig:7a}) and clutter filter ablation (Figure \ref{fig:7b}).

Figure \ref{fig:7a}A shows a 2~mm $\times$ 2~mm window of the mouse brain cortex projected onto a 2D plane, where large arterioles and venules are overlaid with the ground truth capillary mesh used in the simulation. Because the vasculature is interconnected through graph nodes and edges, we performed connectivity analysis on the capillary mesh to identify functionally important capillary nodes using a linear combination of betweenness centrality and PageRank indicators (Figure \ref{fig:7a}B). The combination of these two centrality measures highlights highly influential nodes that are important to capillary network connectivity, indicating their likelihood of providing essential functional information.

This analysis was motivated by the results shown in Figures \ref{fig:7a}C and D, which demonstrate that populating 80--90\% of the large vasculature with high flow (red vessels in Figure \ref{fig:7a}A) requires 3~minutes at the highest microbubble concentrations, whereas the same time period can only populate 50\% of the capillary mesh. This finding agrees with other estimates of the time required to populate the capillary mesh with microbubbles \cite{hingot2021measuring,lowerison2020vivo}. This raises the question: "How many capillary tracks are needed to sufficiently sample capillary function and perfusion in the brain?"

To address this question, we assigned a score to each track passing through the highly important capillary nodes identified in Figure \ref{fig:7a}B, weighted by node importance, where a score of 1 indicates that all highlighted nodes were populated. Figure \ref{fig:7a}E shows that 95\% of highly influential nodes can be populated within 2--3~minutes at the highest concentration of simulated microbubbles. More realistically, accounting for the practical limitation by need of separable microbubbles, lower concentrations (300--500~MB/mm$^3$) may be used, reaching about 80\% coverage in 3~minutes. This evidence suggests that a 3--5~minute scan time may be sufficient for sampling capillary function in an individual, provided that capillary tracks can be confidently and accurately measured.

Secondly, we illustrate the added benefit of The-Bodega simulation framework by investigating how SVD filtering and skull clutter can lead to ablation of slow-moving microbubbles and their corresponding capillary tracks. The reconstructed ULM simulation image with directional overlay is shown in Figure \ref{fig:7b}A, with the zoom box indicated in Figures \ref{fig:7b}B--E. To fairly compare SVD filtering effects on ULM, we compared beamformed and filtered data with combined microbubble, skull clutter, and different eigenvalue cutoffs to the beamformed microbubble-only tracks with no added skull clutter. 

Figure \ref{fig:7b}B illustrates microbubble tracks with no skull clutter, with arrows indicating representative capillary trajectories. In comparison, Figure \ref{fig:7b}C shows that these tracks are ablated after removing 25 of 500 eigenvalues. Additionally, vessels with low track numbers appear noisier, obscuring capillary vessels. To highlight these differences, SVD-filtered ULM images and MB-only ULM images were overlaid (Figure \ref{fig:7b}D). These differences can be cross-referenced with the corresponding velocity map (Figure \ref{fig:7b}E) to demonstrate that these ablated vessels correspond to areas of low-velocity. 

To quantify these effects, we performed Dice, Jaccard index, sensitivity, and specificity analyses on two SVD filtering conditions compared to the MB-only image (Figure \ref{fig:7b}F). We observe that in both conditions, specificity remains high and there is only a slight decrease in sensitivity at higher eigenvalue cutoffs. However, the Dice score and Jaccard index show larger decreases at higher cutoffs, indicating that while ULM tracking can still reconstruct the vascular network (particularly the larger vasculature), the localizations are less precise, especially when reconstructing the capillary mesh network. 

Finally, Figure \ref{fig:7b}G demonstrates the Fourier ring correlation (FRC) measurement of resolution for MB-only, 5-eigenvalue, and 25-eigenvalue cutoffs. The half-bit threshold values show that the resultant resolutions are 9.7~$\mu$m ($\lambda/10$), 14~$\mu$m ($\lambda/7$), and 24~$\mu$m ($\lambda/4$), respectively. This indicates that the loss of capillary vessels diminishes ULM resolution.

\subsection{Effect of motion on heart}

\begin{figure}[!h]
    \centering
    \includegraphics[width=0.9\linewidth]{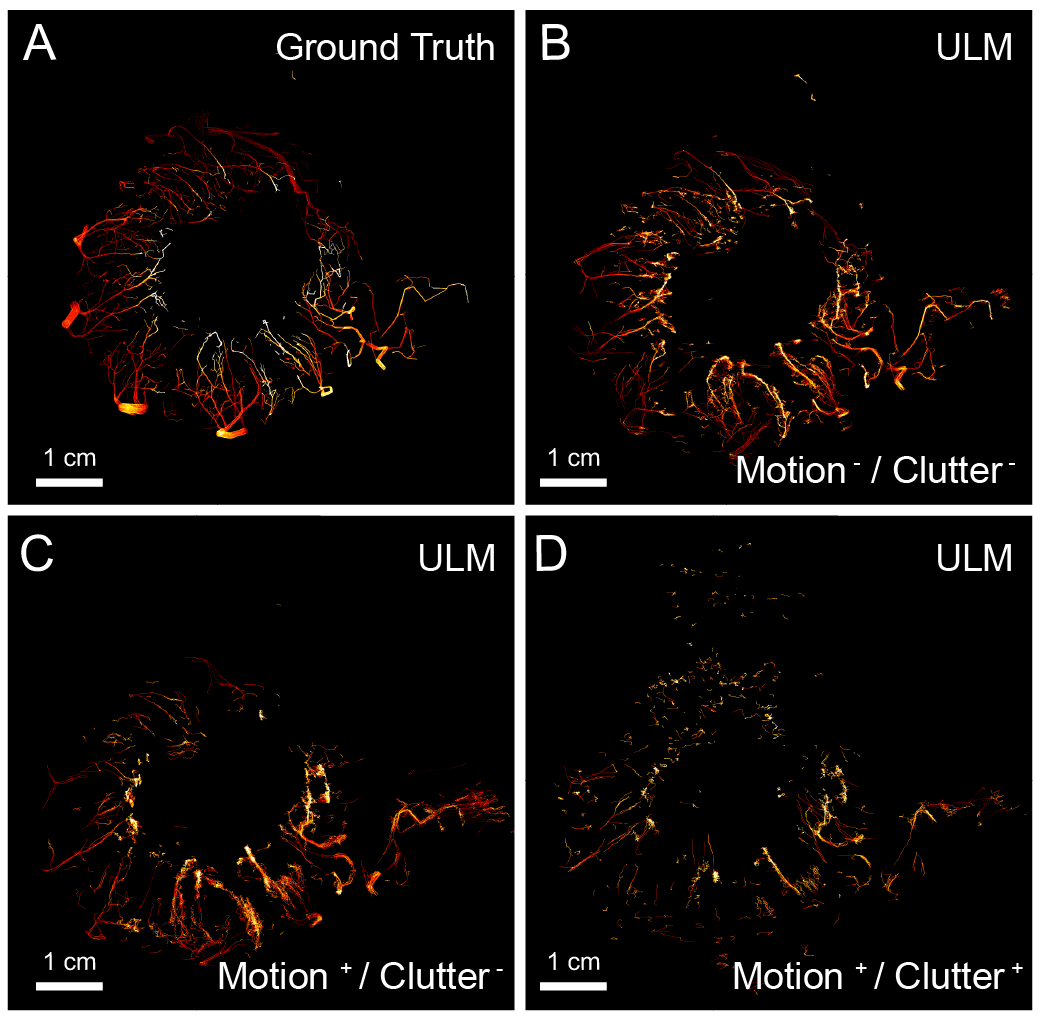}
    \caption{Impact of motion on representative ULM maps. (A) Ground truth representative maps of microbubbles simulated during 18 seconds of acquisition. (B) ULM density maps obtained with clutter when no motion is applied and SVD is performed. (C) ULM density maps of microbubbles obtained when motion is applied but no clutter is present. (D) ULM density maps where both motion and clutter are simulated and SVD is performed.}
    \label{fig:cardiac_results}
\end{figure}

Using The-Bodega, we can independently analyze the effects of motion and tissue clutter on ULM (Figure~\ref{fig:cardiac_results}). Figure~\ref{fig:cardiac_results}A shows the ground truth simulated microbubble positions as they flow through the short-axis microvasculature without any applied motion. These microbubble trajectories are derived directly from the output of the simulation without ultrasound simulation, indicating the best-case scenario. Figure~\ref{fig:cardiac_results}B presents the corresponding ULM density map obtained using standard SVD filtering in the absence of motion. Here, we can see that the vast majority of vessels have been accurately reconstructed. Whereas, Figure~\ref{fig:cardiac_results}C shows the result when motion is applied only to the microbubbles (no tissue clutter), and motion correction is inversely applied to offer a direct comparison to the static case and ground truth cases. Here, we see a greater loss of vasculature, however the larger vasculature can still be recovered and identified. Finally, Figure~\ref{fig:cardiac_results}D displays the ULM density map when both tissue motion and microbubble motion have been applied in the simulated data. After applying clutter filtering and motion correction, some vessels can still be distinguished, but the vast majority are lost during the ULM procedure. Additionally, ULM here also generates false tracks and artifacts. This case highlights the compounding effect of both tissue clutter and motion on ULM as well as demonstrates the difficulty of effective clutter filter and motion corrections algorithms in dynamic imaging conditions.

\subsection{Simulating the neurovascular response for functional ULM}

\begin{figure}[h]
    \centering
    \includegraphics[width=\linewidth]{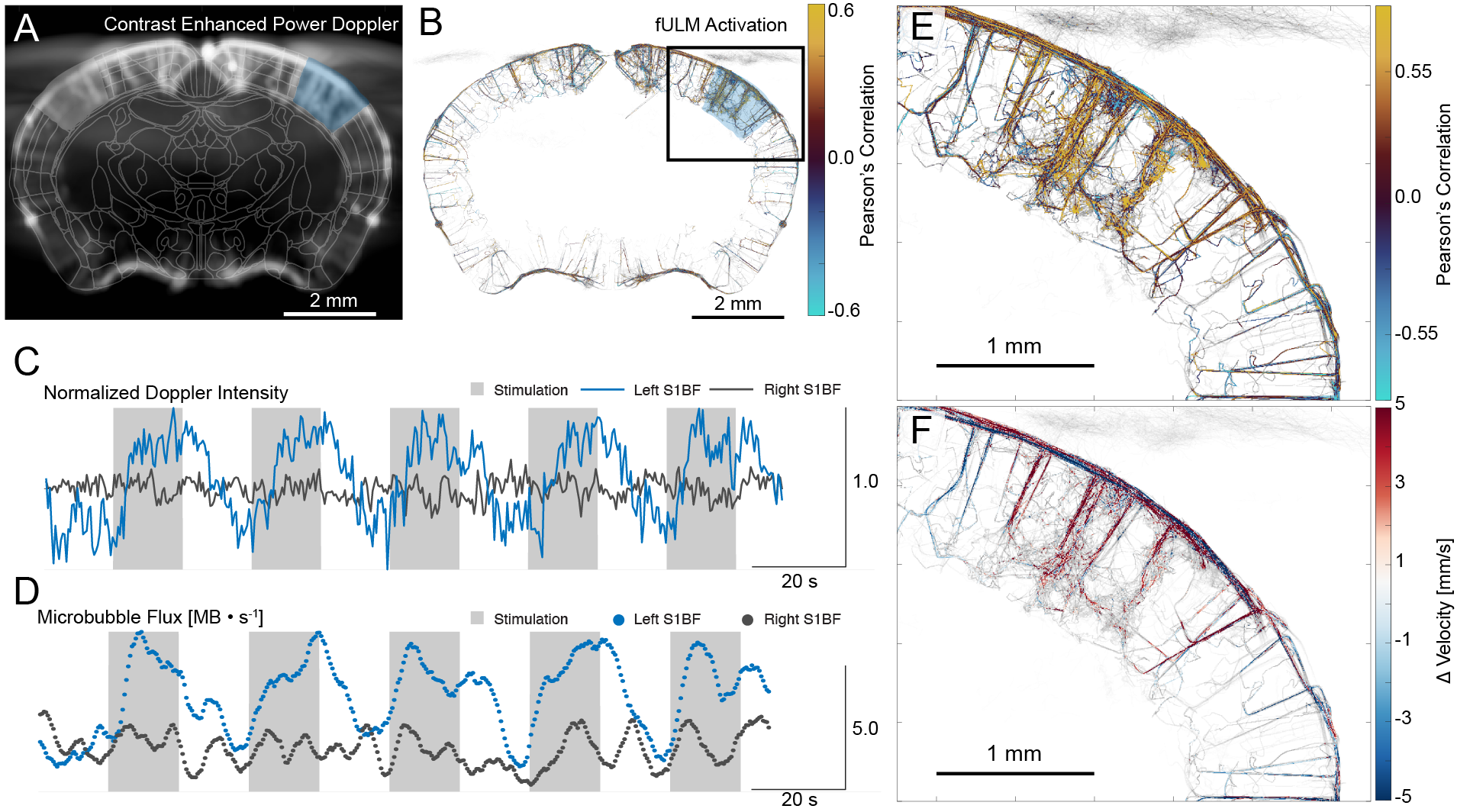}
    \caption{Functional ultrasound localization microscopy reveals simulated neurovascular responses. (A) Contrast-enhanced power Doppler summated over 150~seconds of data and overlaid with the Allen Brain Atlas (plate 250). (B) fULM activation map for the 150~s pulse train. (C) Power Doppler intensity on ipsi- and contralateral S1BF cortices in panel A. (D) Microbubble count and density changes in the corresponding regions in panel B. (E) Zoomed image of activation map in the left barrel field cortex. (F) Zoomed image of velocity changes from baseline at the peak response time point.}
    \label{fig:8}
\end{figure}

Finally, to demonstrate possible applications for dynamic microbubble simulations of microvasculature, we simulated the neurovascular response to whisker stimulation in the left somatosensory barrel field cortex (S1BF). Figure \ref{fig:8}A shows the Allen Brain Atlas coordinate framework overlaid onto the contrast-enhanced power Doppler (summation across all data). The two S1BF ROIs are depicted in blue and gray. In contrast, the resultant ULM map of the activation responses to changes in blood flow and blood volume are illustrated in Figure \ref{fig:8}B. 

The changes in power Doppler intensity (Figure \ref{fig:8}C) and the number of microbubbles flowing through the left S1BF (Figure \ref{fig:8}D) follow the stimulation pulse train injected into the simulation, with relatively stable changes in the contralateral hemisphere. Both the power Doppler and microbubble count changes show strong alignment.

Zooming into the left S1BF region (Figures \ref{fig:8}E and F), we observe the ability to track functional and microbubble speed changes. Using the protocols outlined in \cite{renaudin2022functional}, we calculated the Pearson correlation coefficient between the ULM images (sliding window size of 4~s, 0.4~s stride) and the pulse train (Figure \ref{fig:5}C). We observe positive correlation aggregation within the barrel field region and the corresponding pial vessels that supply this region. Similarly, we observe distributed flow changes (Figure \ref{fig:8}F) in the same penetrating arterioles with large activations. The correlation values reported in this simulation are within the ranges observed \textit{in vivo} \cite{renaudin2022functional}.

\section{Discussion}

Here, we present The-Bodega, a simulation framework for simulating dynamic microvascular environments in the mouse brain cortex and human heart. This simulation package was designed for user exploration and experimentation by implementing all open-source code in a single programming environment with flexible data file formats. We demonstrate that we can not only simulate microbubbles in realistic microvascular networks generated and validated \textit{in vivo}, but also simulate hemodynamics that mimic \textit{in vivo} imaging situations. This dynamic component of ULM simulations not only allow us to answer critical questions in ULM imaging but also supports current research directions in functional, structural, and motion compensated techniques.

For example, we show how realistic heart simulations can be created by injecting movement to visualize microbubble flow in systole and diastole. We demonstrate how the combination of both tissue clutter and beating heart motion corrupts ULM visualization and analysis. An important observation is that motion alone (as in Figure~\ref{fig:cardiac_results}C) does not result in a substantial loss of vascular details. Most vessels can be reconstructed and remain visible after proper motion correction. In contrast, when both motion and tissue clutter are present (Figure~\ref{fig:cardiac_results}D), a large number of vessels are ablated. This suggests that the primary challenge for ULM in moving organs lies not in the motion itself, but in the difficulty of effectively filtering tissue clutter in dynamic conditions. Thus, our simulation framework enables controlled studies in which motion and tissue clutter can be independently manipulated, providing a powerful tool to assess their respective impacts and further develop motion compensation methods for cardiac ULM.

This observation aligns with recent developments in the field, where clutter filtering strategies have been refined to account for spatial variability in tissue motion. For example, adaptive SVD can be applied on windows where tissue motion is similar \cite{Zhang_2022,11050995}. Nonlinear imaging strategies, such as amplitude modulation, have also been adopted with the goal of suppressing the linear tissue response while enhancing the nonlinear signal from microbubbles. This approach aims to reduce reliance on post-processing clutter filtering by physically separating the signals at the acquisition stage \cite{yan2024transthoracic}. While the current work focuses on linear imaging, the simulation pipeline was designed to be modular and flexible, allowing integration with other ultrasound simulation methods, such as nonlinear frameworks \cite{blanken2024proteus}. 

We then illustrate how microvascular whole-brain cortical networks reveal intricate capillary dynamics that are often lost \textit{in vivo}. We elucidate that the long transit times and slow flows across capillaries have large overlaps with the tissue motion space. This is exacerbated by the use of spatiotemporal filters, such as SVD, to remove tissue signal from flowing microbubbles, which ablates these signals and leads to missing recovery of capillary networks, especially across the skull. Thus, improvements to spatiotemporal filtering, reduction of the usage of SVD, or nonlinear imaging modalities \cite{heiles2025nonlinear} may improve the reconstruction of small capillary networks. Moreover, we calculate the saturation time for populating the capillary cortical network through our simulations, indicating that upwards of 30~minutes are required to sample the at least 60\% of the capillary mesh at the highest concentration explored in this study but about 5~minutes are required to sufficiently sample capillary networks function. The limitation in this analysis is that despite simulating 1 million separate microbubbles, not all of these microbubble reach every capillary available in the network due to the stochastic size distribution that may have microbubbles that cannot pass through the each capillary.

The neurovascular response simulations offer an alternative dynamic configuration for The-Bodega simulation - allowing for alterations to the microvascular graphs in any brain region, enabling further investigation into the role that hemodynamics play in neuronal regulation and activity. The simulation methodology (particularly regarding the sampling between two microbubble distributions) is relatively simple and may not entirely reflect microvascular flow changes \textit{in vivo}. However, it serves to facilitate exploration into the combination of neuronal inputs and microvascular flow changes, and the results are in accordance with previous fULM studies \cite{renaudin2022functional}. Moreover, it provides a sandbox for researchers to simulate \textit{in vivo} neurovascular responses and generate or refine the imaging modality to record from a particular target or phenomenon. Thus, it becomes possible to integrate neuronal recordings and simulate the associated neurovascular changes or to use the neurovascular response from functional ultrasound or fMRI to decode the neural and/or immune response. Furthermore, using \textit{in vivo} recordings in this manner may provide a powerful tool for precision medicine.

The-Bodega is also versatile as a deep learning tool as AI and deep learning become more integrated in medical imaging. The flexible and open-source code allows for the generation of any number and combination of synthetic microvascular microbubble and ultrasound data that can be used to train large deep learning models. We envision that researchers will be able to quickly generate their own data for deep learning purposes and develop new data augmentation techniques within the training data itself rather than applying transformations.

As with any simulation framework, there are significant limitations and assumptions inherent to The-Bodega. First, the microbubble simulation results depend on the flow calculated from the input microvascular graph. This means that the flow should be validated before simulation for accurate representation, or Murray's law can be used to relate the expected flow to the radius of the vasculature with less accuracy. Second, the resultant data and input graphs can be large, especially when simulating minutes of high frame rate microbubble data. Thus, the use of fast drives, such as NVMe SSDs, will allow for more efficient transfer and simulation speeds. Additionally, while the ability to simulate in serial is possible, this is not recommended when using large graphs with millions of nodes and edges. The use of parallel processing increases computational efficiency by at least threefold. The framework is also written flexibly such that microbubble and ultrasound simulations can take place as separate jobs in a computing cluster, effectively reducing the computational time further. However, by far the largest computational burden is the ultrasound simulation for simulating highly dense environments like the mouse skull. While SIMUS is implemented on GPU, every simulation performed is in 3D to visualize in- and out-of-plane flow. Additionally, the simulator is linear, and other more complex simulators, such as PROTEUS \cite{blanken2024proteus}, can be used while nonlinear versions of SIMUS are in development.

Finally, this simulation framework is complementary to current ULM simulation modules \cite{heiles2022performance,blanken2024proteus,lerendegui2022bubble}. Specifically, \cite{heiles2022performance} and \cite{lerendegui2022bubble} provided new benchmarking criteria for ULM that have proven effective in guiding the field toward improved ULM imaging. Meanwhile, \cite{blanken2024proteus} introduces novel nonlinear simulators for investigating nonlinear harmonic and pulse inversion techniques. Our simulation framework remains modular, allowing the same microbubble datasets simulated on realistic vasculature to be used as inputs for any ultrasonic simulator (k-Wave \cite{treeby2010k}, Field II \cite{jensen2004simulation}, FULLWAVE \cite{pinton2021fullwave}). The integrated SIMUS ultrasound simulator enables rapid prototyping, while every component of the simulator is written as class functions that are open for alterations and interchangeable. For example, pulsatility calculations can be performed with any algorithm, and real \textit{in vivo} Doppler waveforms can be incorporated. For example, pathological Doppler waveforms can serve as simulation inputs, and the resultant ULM dynamics can be used to identify pathological vascular regions or assess effects on the microvasculature. In summary, the benefit of The-Bodega lies in its ability to simulate the flow of thousands of microbubbles at high sampling rates over minutes in dynamic environments—both cardiovascular and neurovascular—reflecting how ultrasound localization microscopy is currently performed and producing datasets for algorithm, imaging, and deep learning development.

%\appendix
%\section{Example Appendix Section}
%\label{app1}


%%       \cite{<label>} ==> [1]

%%

\bibliographystyle{elsarticle-num} 
\bibliography{bibliography}



%\begin{thebibliography}{00}

%% For numbered reference style

%\bibitem{lamport94}
%  \textit{\LaTeX: a document preparation system},
%  Addison Wesley, Massachusetts,

%\end{thebibliography}
\end{document}